\documentclass[ aps,
                prl,
                twocolumn,
                superscriptaddress
                ]{revtex4}
\usepackage{graphicx}
\usepackage{ulem}
\usepackage[dvipsnames]{xcolor}
\usepackage{amsmath}
\usepackage{amssymb}
\usepackage[colorlinks=true,urlcolor=blue,citecolor=blue,linkcolor=magenta]{hyperref}
\usepackage{braket}

\begin{document}
\title{
Experimental certification of nonclassicality via phase-space inequalities
}

\author{Nicola Biagi}\email{nicola.biagi@ino.cnr.it}
\affiliation{Istituto Nazionale di Ottica (CNR-INO), L.go E. Fermi 6, 50125 Florence, Italy}
\affiliation{LENS and Department of Physics $\&$ Astronomy, University of Firenze, 50019 Sesto Fiorentino, Florence, Italy}
\author{Martin Bohmann}\email{martin.bohmann@oeaw.ac.at}
\affiliation{Institute for Quantum Optics and Quantum Information - IQOQI Vienna, Austrian Academy of Sciences, Boltzmanngasse 3, 1090 Vienna, Austria}
\affiliation{
Vienna Center for Quantum Science and Technology (VCQ), Vienna, Austria}
\author{Elizabeth Agudelo}
\affiliation{Institute for Quantum Optics and Quantum Information - IQOQI Vienna, Austrian Academy of Sciences, Boltzmanngasse 3, 1090 Vienna, Austria}
\author{Marco Bellini}
\affiliation{Istituto Nazionale di Ottica (CNR-INO), L.go E. Fermi 6, 50125 Florence, Italy}
\affiliation{LENS and Department of Physics $\&$ Astronomy, University of Firenze, 50019 Sesto Fiorentino, Florence, Italy}
\author{Alessandro Zavatta}
\affiliation{Istituto Nazionale di Ottica (CNR-INO), L.go E. Fermi 6, 50125 Florence, Italy}
\affiliation{LENS and Department of Physics $\&$ Astronomy, University of Firenze, 50019 Sesto Fiorentino, Florence, Italy}

\begin{abstract}
    In spite of its fundamental importance in quantum science and technology, the experimental certification of nonclassicality is still a challenging task, especially in realistic scenarios where losses and noise imbue the system.
    Here, we present the first experimental implementation of the recently introduced phase-space inequalities for nonclassicality certification, which conceptually unite phase-space representations with correlation conditions.
    We demonstrate the practicality and sensitivity of this approach by studying nonclassicality of a family of noisy and lossy quantum states of light.
    To this end, we experimentally generate single-photon-added thermal states with various thermal mean photon numbers and detect them at different loss levels.
    Based on the reconstructed Wigner and Husimi $Q$ functions, the inequality conditions detect nonclassicality despite the fact that the involved distributions are nonnegative, which includes cases of high losses ($93$\%) and cases where other established methods do not reveal nonclassicality.
    We show the advantages of the implemented approach and discuss possible extensions that assure a wide applicability for quantum science and technologies.
\end{abstract}
\date{\today}
\maketitle

\paragraph{Introduction.---}	
    Nonclassicality relates to the inability to describe certain effects or physical systems by a classical theory.
    It represents the basis for quantum technologies.
    Therefore, it is of fundamental scientific and technological importance to detect and certify such nonclassical quantum effects.
    A multiplicity of different approaches and methods to verify all different notions of quantumness have been introduced.
    However, the certification of nonclassical features in realistic, complex, and noisy systems is still a challenging task.
    
    The phase-space formalism, that is based on the description of physical systems and their dynamics through phase-space distributions, is an elegant and widely used method that allows us to characterize classical and quantum systems on the same footing \cite{S01,ZFC05}.
    In order to express quantum states through phase-space distributions, the concept has been extended to the so-called quasiprobability distributions, which may attain negative values (see \cite{SW18,SV20} for a detailed introduction).
    Such negativities constitute a direct signature of nonclassicality.
    The quantum-state representation by phase-space quasidistributions is, however, not unique.
    The best known and most used quasidistributions are the Wigner function \cite{W32}, the Husimi $Q$ function \cite{H40}, and the Glauber-Sudarshan $P$ function \cite{S63,G63}.
    In quantum optics, negativities in the latter are the base of the very definition of nonclassicality \cite{TG65,M86}. 
    However, the $P$ function is in most cases a highly singular distribution that cannot be reconstructed experimentally, except for some notable exceptions \cite{KVPZB08}.
    Therefore, in many cases one rather considers the Wigner or $Q$ functions, which can be directly measured \cite{WV96,LMKMIW96,BRWK99,LCGS10} or reconstructed via homodyne-detection techniques \cite{WVO99,LR09}.
    Expressing quantum states through phase-space distributions other than the $P$ function has the drawback that not all nonclassical features will be recognized, since often they turn out to be nonnegative distributions for nonclassical states.
    For example, nonclassical Gaussian states, such as squeezed states, are represented by nonnegative Wigner functions \cite{WPGCRSL12}, and nonclassicality cannot be directly inferred from $Q$ functions as they are nonnegative by definition.
    
    Among the multitude of quantum states of light, the single-photon state has a paramount relevance.
    On the one hand, it is the state of the fundamental excitation of the quantized light field, representing its particle character.
    On the other hand, single photons are fundamental carriers of information and are widely used for quantum information applications \cite{KMNRDM07,OFJ09,KMSUZ16}.
    Such states are realized in a plethora of experimental platforms \cite{MBBMBKK04,AET16,BLCB16,KK19,LTCMLKT19} and their phase-space functions are routinely reconstructed; cf., e.g., \cite{LS02,BTBSSV18,BBFPT19,SPBTEWLNLGVASW20}.
    Unfortunately, the certification of nonclassicality of single-photon states can be a difficult task when losses and noise impair the system.
    In particular, the Wigner function of a lossy single-photon state is nonnegative for losses above $50$\% \cite{LHABMS01}.
    The situation is more difficult when additional noise contributions have to be considered.
    A special example of such noisy states is the family of single-photon-added thermal states (SPATS) \cite{AT92,ZPB07,BZ10}:
    $\hat\rho=\mathcal{N}\hat a^\dagger\hat\rho_{\mathrm{th}}\hat a$, with $\mathcal{N}$ being its normalization constant, $\hat a$ and $\hat a^\dagger$ the annihilation and creation operators, respectively, and $\hat\rho_{\mathrm{th}}=1/(\bar{n}+1)\sum_{k=0}^\infty \left[\bar{n}/(\bar{n}+1)\right]^k |k\rangle\langle k|$ a thermal state.
    Including losses, SPATS are characterized by a two-parameter state space that allows us to investigate different regimes.
    In particular, they feature parameter regions in which the state's nonclassicality cannot be certified by the established conditions based on phase-space distributions and the photon-number distribution \cite{ZPB07,KV14}.
    This illustrates the need for the development and implementation of experimentally accessible, sensitive, and noise-robust nonclassicality tests.
    
    In this letter, we present the first experimental implementation of the recently introduced phase-space-distribution inequalities \cite{BA20,BAS20} to certify the nonclassicality of SPATS.
    We generate a family of SPATS with different thermal mean photon numbers~($\bar{n}$) and measure them with balanced homodyne detection at different detection efficiencies~($\eta$).
    For the implementation of the nonclassicality inequality conditions, we reconstruct the Wigner and Husimi $Q$ functions of the states via quantum-state tomography.
    We are able to certify nonclassicality despite the fact that the obtained phase-space distributions are nonnegative.
    Importantly, the phase-space conditions reveal nonclassicality for efficiencies as low as $7\%$ and in parameter regions where other nonclassicality tests fail to do so.
    Thus, we demonstrate that phase-space inequalities are powerful certification tools that go far beyond nonclassicality detection based on the sheer value of phase-space distributions.
    Our results pave the way for the efficient and reliable characterisation of quantum states under the influence of loss and noise being ever more important in the context of developing practical quantum technologies.
    
\paragraph{Phase-space distribution inequalities.---}
    Only very recently phase-space-distribution inequalities for testing nonclassicality were derived \cite{BA20,BAS20,PLN20}.
	This approach allows one to certify the nonclassicality of a given state by considering different points of the same distribution or by relating different phase-space distributions evaluated in the same point in phase space with each other.
	Importantly, these phase-space conditions provide the possibility of revealing nonclassicality even if the involved phase-space distributions are nonnegative and, thus, drastically extending the certification capability based on such quasidistributions. Furthermore, being based on correlations between quasidistributions at different points, a full tomographic reconstruction can in principle be avoided by using direct sampling measurements \cite{WV96,LMKMIW96,BRWK99,LCGS10}.
	
	To illustrate the strength and practicality of these phase-space inequalities, we will use two particular nonclassicality conditions for the certification of experimentally generated lossy SPATS.
	The first condition is based on the matrix of phase-space distributions \cite{BAS20} and relates the Wigner function in different points in phase space ($\alpha_1\neq\alpha_2$) with each other
	\begin{align}\label{eq:WW}
		W(\alpha_1)W(\alpha_2)-e^{-|\alpha_2-\alpha_1|^2}W[(\alpha_1+\alpha_2)/2]^2< 0.
	\end{align}
	Detecting classically unattainable fine features of the Wigner function, such a nonclassicality condition is related to the so-called sub-Planck phase-space structures \cite{Z01}.
	The second one includes the Wigner and $Q$ functions evaluated in the same phase-space point $\alpha$ \cite{BA20}:
	\begin{align}\label{eq:WQ}
		W(\alpha)-2\pi\, Q(\alpha)^2< 0.
	\end{align}
	In this case, the detected nonclassicality is related to the influence of different operator ordering associated to the two different phase-space distributions \cite{CG69}.
	As we will show, these conditions are easy to implement and will allow us to certify nonclassicality of SPATSs in regimes where other well-established methods fail to do so.

\begin{figure}[t]
 		\centering
 		\includegraphics[width=\columnwidth]{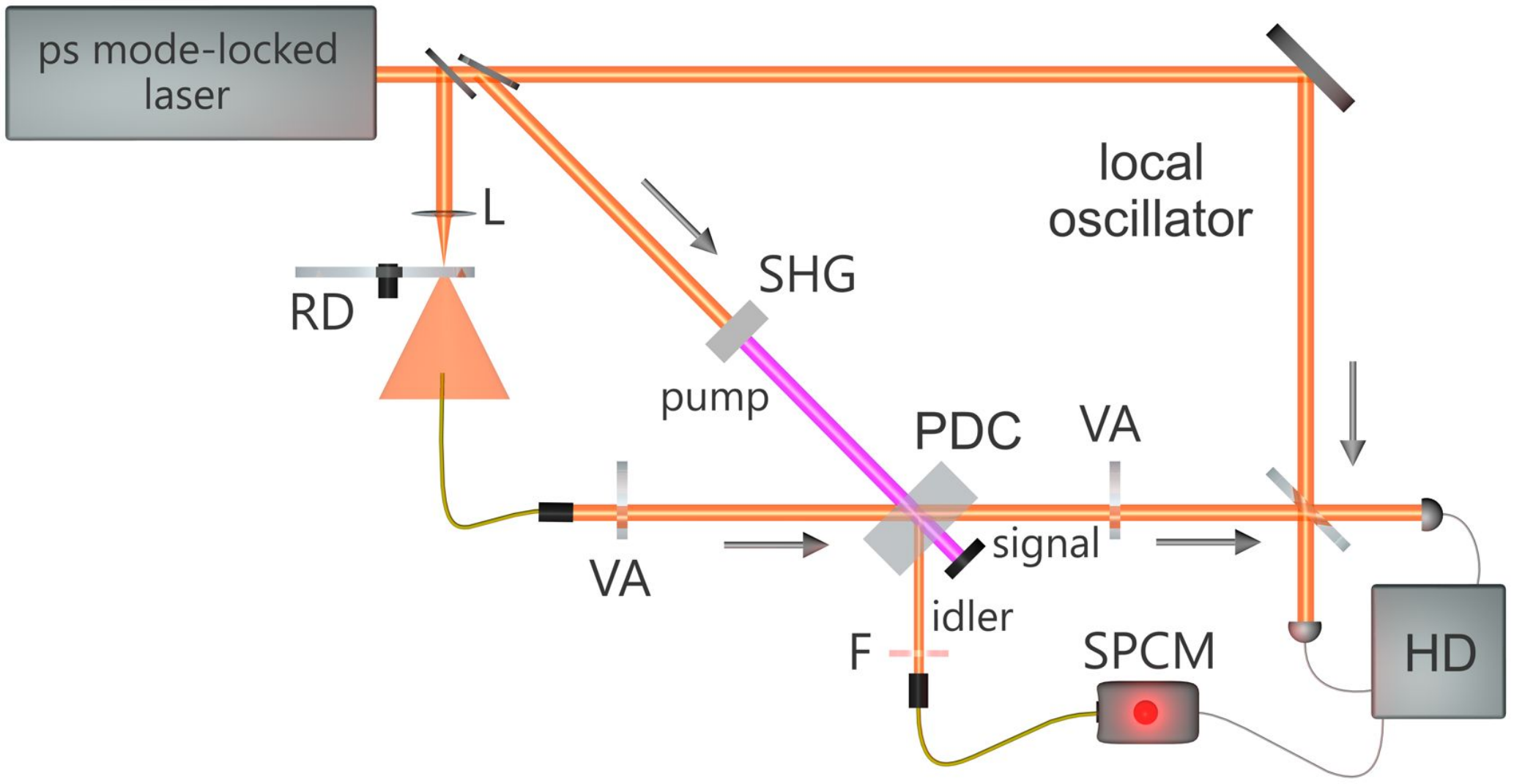}
 		\caption{
 		        Experimental setup for the heralded generation and detection of single-photon-added thermal states of light (SPATS).
 		        See the text for details.
 			}
 		\label{fig:ExpSetup}
 	\end{figure}

\begin{figure*}[t]
 		\centering
 		\includegraphics[width=\textwidth]{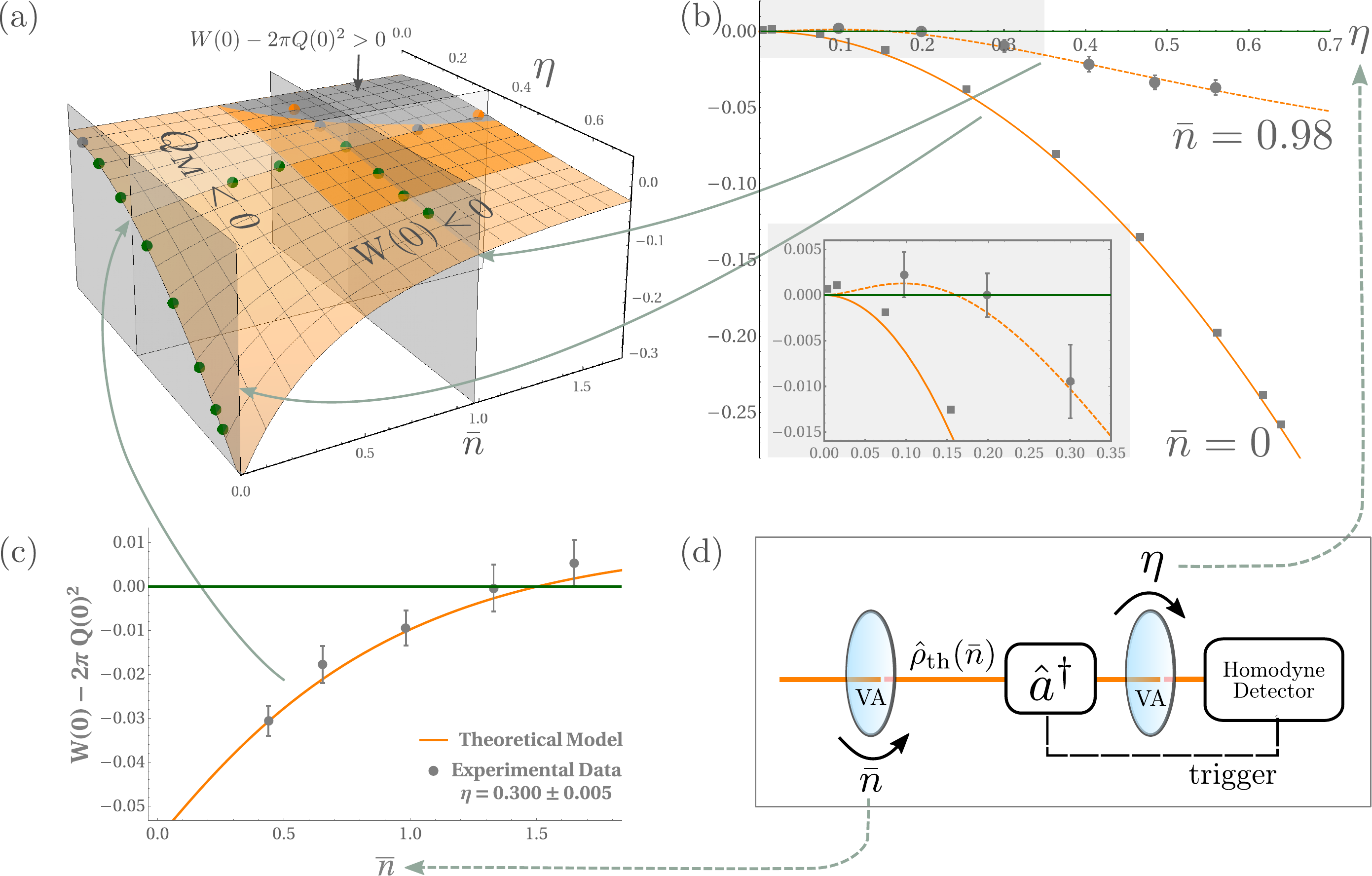}
 		\caption{
 		Evaluation of single-point nonclassicality condition \eqref{eq:WQ}.
 		(a) Different states (points) are depicted in the two-dimensional $\bar{n}$-$\eta$-space.
 		The orange surface indicates the parameter region where nonclassicality can be detected via condition \eqref{eq:WQ}.
 		(b) $\bar{n}=0$ and $0.98$ cross-sections.
 		(c) $\eta=0.3$ cross-section.
 		(d) Illustration of the experiment and the parameter control.
        Vertical error bars correspond to the standard deviation of the obtained values. 
 			}
 		\label{fig:SPWRedults}
 	\end{figure*}

\paragraph{Experiment.---}
    Our experimental setup is based on a mode-locked Ti:sapphire laser emitting 1.5 ps long pulses at 786 nm, with a repetition rate of 81 MHz (see Fig. \ref{fig:ExpSetup}).
    Its output is split in three main parts.
    We produce a pseudo-thermal state of light by using a lens (L) to focus the first part of the main laser beam onto a rotating ground glass disk (RD)\cite{A65}.
    Collecting a small portion of the light scattered by the disk into a single-mode fiber, we obtain a thermal state in a single spatial mode, which is then aligned along the input signal mode of a parametric down-conversion (PDC) crystal.
    The second portion of the beam is frequency-doubled at a second-harmonic generation (SHG) stage and provides the pump for the frequency-degenerate non-collinear PDC process.
    By conditioning on a click in the single-photon counting module (SPCM) of the output idler mode after tight spatial and spectral filtering (F), we implement a single-photon addition operation ($\hat{a}^\dagger$) onto the signal mode, resulting in the desired SPATS.
    The third portion of the laser output serves as the local oscillator (LO) for a time-domain balanced homodyne detector (HD) used to acquire quadrature data for the tomographic reconstruction of the Wigner and $Q$ functions of the quantum state in the signal mode.
    Two variable attenuators (VA), placed along the signal path before and after the PDC crystal, are used to control the thermal mean photon number ($\bar{n}$) and the detection efficiency ($\eta$) with which the state is recorded, respectively.
    See Fig. \ref{fig:SPWRedults}(d) for a pictorial representation of the setup.

    To experimentally test conditions \eqref{eq:WW} and \eqref{eq:WQ}, we acquired about $260$k phase-randomized quadrature values for different settings of $\bar{n}$ (including the single-photon case, $\bar{n}=0$) and $\eta$.
    The $\eta$ and $\bar{n}$ parameters, and their respective errors were estimated by fitting the single-photon and SPATS quadrature data as function of the two VA settings.
    An iterative maximum likelihood procedure was used to reconstruct the density matrix associated to each data set and to subsequently calculate their respective Wigner and $Q$ functions \cite{HRFJ04,L04}.
    Statistical errors are estimated via a bootstrap method based on 50 resampled quadrature data sets.

\paragraph{Results.---}
    We use the phase-space inequalities to certify nonclassicality of our experimentally generated SPATS.
    To benchmark our results, we compare the performance to two well-established and widely-used nonclassicality tests: the Mandel $Q_{\mathrm{M}}$ parameter \cite{M79} and the negativities of the Wigner function.
    Note that these two approaches are based on very different aspects of quantum states, i.e., the photon-number distribution and a phase-space representation.
    Furthermore, it is known that the Wigner function of any SPATS shows negativities for $\eta>50\%$ independent of $\bar{n}$, and that the Mandel $Q_{\mathrm{M}}$ detects nonclassicality for $\bar{n}<1/\sqrt{2}$ for any nontrivial loss $\eta>0$ \cite{KV14}.
    Hence, these two ways of certifying nonclassicality ideally complement each other in the sense that one is independent of $\eta$ and the other of $\bar{n}$.
    Therefore, they provide an excellent basis of comparison for the phase-space inequalities, which allows to assess their performance.
    
    \begin{figure}[t]
 		\includegraphics[width=\columnwidth]{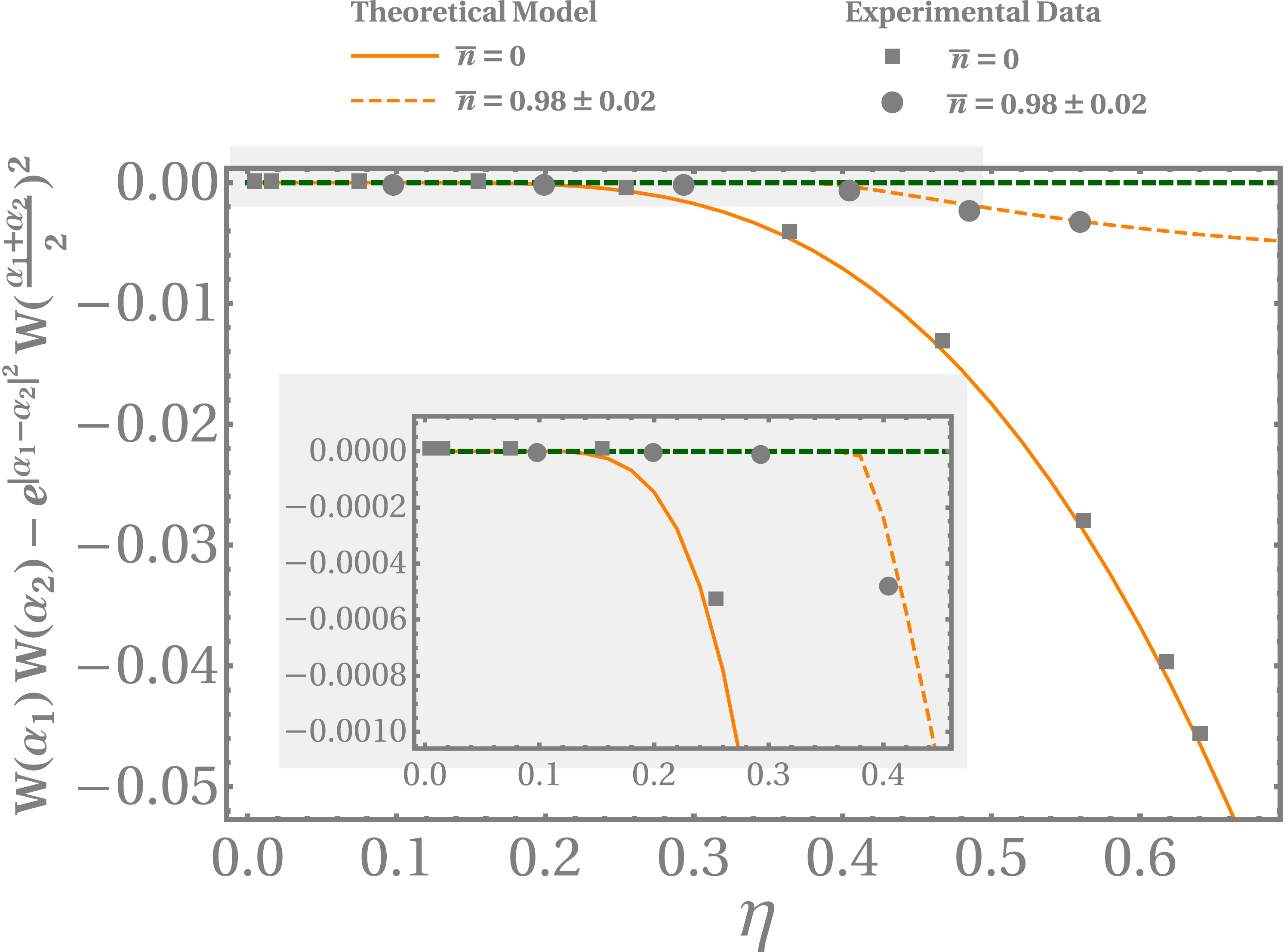}
 		\caption{
 		Evaluation of the multi-point nonclassicality condition \eqref{eq:WW} in dependence on the detection efficiency ($\eta$), for the states with thermal mean photon numbers (squares and orange solid line) $\bar{n}=0$ and (circles and orange dashed line) $\bar{n}=0.98\pm 0.02$.
 		Error bars are not visible since they are smaller than the points area.
 		}
 		\label{fig:MPWRedults}
 	\end{figure}
    
    We start our nonclassicality analysis of the SPATS based on condition \eqref{eq:WW}, which only requires the knowledge of the Wigner function of the state.
    The results are presented in Fig.~\ref{fig:MPWRedults}, where we show the dependence on the efficiency $\eta$ for two thermal mean photon numbers $\bar{n}$.
    For evaluating the multi-point condition \eqref{eq:WW}, we numerically estimate the phase-space points that optimize it for each parameter pair $(\eta,\bar{n})$ and calculate its value from the experimentally reconstructed Wigner function at those points.
    In the single-photon case ($\bar{n}=0$), it clearly reveals nonclassicality down to $\eta\approx25$\% where the Wigner function is long nonnegative.
    Hence, by simply comparing different points of the Wigner function, it is possible to certify nonclassicality despite the fact that the distribution is nonnegative.
    Note that this approach does not require any additional analysis or calculations as it solely relates different points of the already reconstructed Wigner function with each other.
    In this sense, Eq. \eqref{eq:WW} provides a clever and efficient way of detecting nonclassicality based on the available information of the quantum state.
    
    In the case of considerably strong thermal background $(\bar{n}=0.98)$ , it is possible to certify nonclassicality down to efficiencies of about $40\%$; cf. Fig. \ref{fig:MPWRedults}.
    Although, this corresponds to a lower loss threshold compared to the single-photon case, the situation is even more interesting.
    Additionally to possessing a nonnegative Wigner function, the SPATS with $\eta=40\%$ and $\bar{n}=0.98$ also shows a positive Mandel parameter $Q_{\mathrm{M}}$, which does not detect nonclassicality either.
    Thus, this example shows that, besides its simplicity, the phase-space inequality \eqref{eq:WW} certifies nonclassicality in parameter regions inaccessible by these established nonclassicality certifiers.
    
    Figure \ref{fig:SPWRedults} summarizes the experimental results concerning the test of the single-point condition \eqref{eq:WQ} evaluated at the origin of phase space ($\alpha=0$), where a maximum violation is expected for SPATS.
    The results for the different lossy SPATS are depicted (points) in the two-dimensional $(\eta,\bar{n})$ state space in Fig. \ref{fig:SPWRedults}(a).
    The light orange area indicates the region where the nonclassicality of the state can also be detected either by $Q_{\mathrm{M}}<0$ or the negativities of the Wigner function.
    The dark orange area is where the two above conditions fail but our single-point condition \eqref{eq:WQ} still certifies nonclassicality, showing a clear advantage of the phase-space inequalities.
    In Fig. \ref{fig:SPWRedults}(b) and (c), we present different cross sections through the parameter plane, where the solid and dashed lines represent the theoretically expected behavior. 
    The control of the parameters in the experiment is illustrated in Fig. \ref{fig:SPWRedults}(d).
    
    We first explore two cross-sections of the $W(0)-2\pi Q(0)^2$ surface for two fixed values of the thermal mean photon number ($\bar{n}=0$ and $\bar{n}=0.98\pm0.02$) by varying the efficiency $\eta$ in Fig. \ref{fig:SPWRedults}(b).
    For $\bar{n}=0$, we can certify nonclassicality via condition \eqref{eq:WQ} for efficiencies as low as $\eta\approx 7\%$ ($93$\% loss) even considering statistical and experimental errors [see inset in Fig. \ref{fig:SPWRedults}(b)].
    For SPATS with $\bar{n}\approx 1$, we detect nonclassicality down to $\eta \approx 30\%$.
    Note that this cross section includes nonclassical states that are neither detected by the Wigner function nor the Mandel parameter.
    Hence, this case shows that the inequality \eqref{eq:WQ} detects nonclassicality under the influence of strong loss and thermal background.
    
    For the second cross section in Fig. \ref{fig:SPWRedults}(c), we fixed the efficiency at $\eta=0.3$ and investigate the influence of the thermal mean photon number.
    In this case, we can experimentally detect nonclassicality in a wide parameter range up to $\bar{n}\approx1$.
    Increasing the number of collected quadrature data it could be possible to reduce the statistical error and probably it could be possible to push $\bar{n}$  further to $\approx1.5$, i.e., to a regime where the classical thermal contribution dominates the energy of the state.
    
    For all cases analyzed, we observe an excellent agreement between the theoretically expected behaviour (lines) and the experimental data [cf. Figs. \ref{fig:MPWRedults} and \ref{fig:SPWRedults}], which underlines the reliability and robustness of the presented methods.
    Furthermore, the current limits in detecting nonclassicality in the entire orange area in Fig. \ref{fig:SPWRedults}(a) are only due to statistical uncertainties that can be easily reduced by increasing the number of collected quadrature data.
    Additionally, the parameter region for which nonclassicality can be detected could be increased (that is, the orange area of Fig. \ref{fig:SPWRedults}(a) could be in principle extended to the entire parameter space) by using other (higher-order) phase-space conditions \cite{BAS20}.

\paragraph{Discussion and conclusions.---}
    Our experimental results present the first implementation of recently introduced phase-space-inequality conditions for the certification of nonclassicality.
    In this way, we experimentally demonstrated that it is possible to certify nonclassicality of states directly from their nonnegative phase-space distributions.
    We showed that these conditions are particularly useful under unfavorable conditions, where quantum features are obscured by strong losses and background radiation, and where it is genuinely hard to certify them.
    Despite their sensitivity, it is remarkable that the presented inequality conditions are easy to implement by just comparing different values of phase-space distributions with each other, which offers a straightforward and ready implementation without the need of any involved calculations.
    The soundness of our results is further strengthened by excellent agreement of the obtained results with theoretical predictions.
    
    We benchmarked the phase-space-inequality results against two established nonclassicality conditions: negativities in the Wigner function and the Mandel $Q_{\mathrm{M}}$ parameter.
    For the considered lossy SPATS these two conditions are a natural choice of comparison due to their complementary dependence on the loss and thermal noise parameters as explained above.
    We showed that the phase-space inequalities do not just detect nonclassicality of all SPATS that are also detected by Wigner negativities and $Q_{\mathrm{M}}$, but even in parameter regions where these last two fail to certify nonclassicality; cf., e.g., the dark orange region in Fig. \ref{fig:SPWRedults}(a).
    This demonstrates the strength and sensitivity of the phase-space inequalities.
    Additionally, our experimental results show that only one phase-space condition is sufficient to witness nonclassicality in very different parameter regions where otherwise one would need to implement different tests in order to certify nonclassicality.
    Summarizing, these findings illustrate the efficiency and practicality of phase-space inequalities for certifying nonclassicality under challenging noisy conditions.
    
    It is important to highlight that the used approach is universal in the sense that it is applicable to any quantum state including both---continuous and discrete variable states.
    Furthermore, in contrast to intensity-based conditions, the phase-space inequalities offer the possibility of revealing phase-sensitive nonclassical features.
    An extension of the presented analysis to higher-order conditions and the detection of quantum correlations between different modes is also possible \cite{BAS20,BCBZ20}.
    The phase-space inequalities are universally applicable in any experimental scenario where it is possible to retrieve phase-space distributions, either by a full tomographic reconstruction or by direct sampling.
    Furthermore, the method is not limited to photonic systems but can be readily applied to other physical platforms including microwave-cavity \cite{GFPMTMGSD20}, mechanical-oscillators \cite{MLLV19}, and trapped-ion \cite{FH20} experiments.
    Thus, phase-space inequalities will find applications in many different experiments throughout various physical systems.

    N.B., M.B., and A.Z. gratefully acknowledge the support of Ente Cassa di Risparmio di Firenze under the project “MOSTO”and of the EU under the ERA-NET QuantERA project “ShoQC” and the FET Flagship on Quantum Technologies project "Qombs" (grant no. 820419).
    E.A. acknowledges funding from the European Union's Horizon 2020 research and innovation programme under the Marie Sk\l{}odowska-Curie IF (InDiQE - EU project 845486).
	



\begin{thebibliography}{99}

    \bibitem{S01}
		W. P. Schleich,
		\textit{Quantum Optics in Phase Space}
		(Wiley-VCH, Berlin, 2001).
	\bibitem{ZFC05}
		C. Zachos, D. Fairlie, and T. Curtright,
		\textit{Quantum Mechanics in Phase Space}
		(World Scientific, Singapore, 2005).
	\bibitem{SW18}
		J. Sperling and I. A. Walmsley,
		\textit{Quasiprobability representation of quantum coherence},
		\href{https://doi.org/10.1103/PhysRevA.97.062327}{Phys. Rev. A \textbf{97}, 062327 (2018)}.
	\bibitem{SV20}
		J. Sperling and W. Vogel,
		\textit{Quasiprobability distributions for quantum-optical coherence and beyond},
		\href{https://doi.org/10.1088/1402-4896/ab5501}{Phys. Scr. \textbf{95}, 034007 (2020)}.
	\bibitem{W32}
		E. Wigner,
		\textit{On the Quantum Correction For Thermodynamic Equilibrium},
		\href{https://doi.org/10.1103/PhysRev.40.749}{Phys. Rev. \textbf{40}, 749 (1932)}.
	\bibitem{H40}
		K. Husimi,
		\textit{Some formal properties of the density matrix},
		\href{https://doi.org/10.11429/ppmsj1919.22.4_264}{Proc. Phys. Math. Soc. Jpn. \textbf{22}, 264 (1940)}.
	\bibitem{G63}
		R. J. Glauber,
		\textit{Coherent and Incoherent States of the Radiation Field},
		\href{https://doi.org/10.1103/PhysRev.131.2766}{Phys. Rev. \textbf{131}, 2766 (1963)}.
	\bibitem{S63}
		E. C. G. Sudarshan,
		\textit{Equivalence of Semiclassical and Quantum Mechanical Descriptions of Statistical Light Beams},
		\href{https://doi.org/10.1103/PhysRevLett.10.277}{Phys. Rev. Lett. \textbf{10}, 277 (1963)}.
	\bibitem{TG65}
		U. M. Titulaer and R. J. Glauber,
		\textit{Correlation functions for coherent fields}, 
		\href{https://doi.org/10.1103/PhysRev.140.B676}{Phys. Rev. \textbf{140}, B676 (1965)}.
	\bibitem{M86}
		L. Mandel, 
		\textit{Non-classical states of the electromagnetic field},
		\href{https://doi.org/10.1088/0031-8949/1986/T12/005}{Phys. Scr. T \textbf{12}, 34 (1986)}.
	\bibitem{KVPZB08}
		T. Kiesel, W. Vogel, V. Parigi, A. Zavatta, and M. Bellini,
		\textit{Experimental determination of a nonclassical Glauber-Sudarshan P function},
		\href{https://doi.org/10.1103/PhysRevA.78.021804}{Phys. Rev. A \textbf{78}, 021804(R) (2008)}.
	\bibitem{WV96}
    	S. Wallentowitz and W. Vogel,
    	\textit{Unbalanced homodyning for quantum state measurements},
    	\href{https://doi.org/10.1103/PhysRevA.53.4528}{Phys. Rev. A \textbf{53}, 4528 (1996)}.
    \bibitem{LMKMIW96}
	    D. Leibfried, D. M. Meekhof, B. E. King, C. Monroe, W. M. Itano, and D. J. Wineland,
	    \textit{Experimental Determination of the Motional Quantum State of a Trapped Atom},
        \href{https://doi.org/10.1103/PhysRevLett.77.4281}{Phys. Rev. Lett. 77, 4281 (1996)}.
    \bibitem{BRWK99}
    	K. Banaszek, C. Radzewicz, K. W\'odkiewicz, and J. S. Krasi\'{n}ski,
    	\textit{Direct measurement of the Wigner function by photon counting},
    	\href{https://doi.org/10.1103/PhysRevA.60.674}{Phys. Rev. A \textbf{60}, 674 (1999)}.
    \bibitem{LCGS10}
    	K. Laiho, K. N. Cassemiro, D. Gross, and C. Silberhorn,
    	\textit{Probing the Negative Wigner Function of a Pulsed Single Photon Point by Point},
    	\href{https://doi.org/10.1103/PhysRevLett.105.253603}{Phys. Rev. Lett. \textbf{105}, 253603 (2010)}.
	\bibitem{WVO99}
		D.-G. Welsch, W. Vogel, and T. Opatrn\'y,
		\textit{Homodyne detection and quantum-state reconstruction},
		\href{https://doi.org/10.1016/S0079-6638(08)70389-5}{Progress in optics, edited by E. Wolf, Vol. XXXIX (1999), pp. 63-211}.
	\bibitem{LR09}
		A. I. Lvovsky and M. G. Raymer,
		\textit{Continuous-variable optical quantum-state tomography},
		\href{https://doi.org/10.1103/RevModPhys.81.299}{Rev. Mod. Phys. \textbf{81}, 299 (2009)}.
	\bibitem{WPGCRSL12}
		C. Weedbrook, S. Pirandola, R. Garc\'{\i}a-Patr\'on, N. J. Cerf, T. C. Ralph, J. H. Shapiro, and S. Lloyd,
		\textit{Gaussian quantum information},
		\href{https://doi.org/10.1103/RevModPhys.84.621}{Rev. Mod. Phys. \textbf{84}, 621 (2012)}.
	\bibitem{KMNRDM07}
		P. Kok, W. J. Munro, K. Nemoto, T. C. Ralph, J. P. Dowling, and G. J. Milburn,
		\textit{Linear optical quantum computing with photonic qubits},
		\href{https://doi.org/10.1103/RevModPhys.79.135}{Rev. Mod. Phys. \textbf{79}, 135 (2007)}.
	\bibitem{OFJ09}
		J. L. O'Brien, A. Furusawa, and J. Vu\v{c}kovi\'{c},
		\textit{Photonic quantum technologies},
		\href{https://doi.org/10.1038/nphoton.2009.229}{Nat. Photonics \textbf{3}, 687 (2009)}.
	\bibitem{KMSUZ16}
		M. Krenn, M. Malik, T. Scheidl, R. Ursin, and A. Zeilinger,
		\textit{Quantum communication with photons},
		in \textit{Optics in Our Time} (Springer, Cham, 2016), pp. 455--482.
	\bibitem{MBBMBKK04}
		J. McKeever, A. Boca, A. D. Boozer, R. Miller, J. R. Buck, A. Kuzmich, and H. J. Kimble,
		\textit{Deterministic Generation of Single Photons from One Atom Trapped in a Cavity},
		\href{https://doi.org/10.1126/science.1095232}{Science \textbf{303}, 1992 (2004)}
    \bibitem{AET16}
		I. Aharonovich, D. Englund, and M. Toth,
		\textit{Solid-state single-photon emitters},
		\href{https://doi.org/10.1038/nphoton.2016.186}{Nature Photon \textbf{10}, 631 (2016).}
	\bibitem{BLCB16}
		M. Bock, A. Lenhard, C. Chunnilall, and C. Becher,
        \textit{Highly efficient heralded single-photon source for telecom wavelengths based on a PPLN waveguide},
        \href{https://doi.org/10.1364/OE.24.023992}{Opt. Express \textbf{24}, 23992 (2016)}.
    \bibitem{KK19}
        F. Kaneda and P. G. Kwiat,
        \textit{High-efficiency single-photon generation via large-scale active time multiplexing},
        \href{https://doi.org/10.1126/sciadv.aaw8586}{Science Advances \textbf{5}, eaaw8586 (2019)}.
    \bibitem{LTCMLKT19}
        P. Lombardi, M. Trapuzzano, M. Colautti, G. Margheri, M. L\'opez, S. K\"uck, C. Toninelli
        \textit{A Molecule-Based Single-Photon Source Appliedin Quantum Radiometry},
        \href{https://arxiv.org/abs/1908.00616}{arXiv:1908.00616 [quant-ph]}
	\bibitem{LS02}
		A. I. Lvovsky and J. H. Shapiro,
		\textit{Nonclassical character of statistical mixtures of the single-photon and vacuum optical states},
		\href{https://doi.org/10.1103/PhysRevA.65.033830}{Phys. Rev. A \textbf{65}, 033830 (2002)}.
	\bibitem{BTBSSV18}
    	M. Bohmann, J. Tiedau, T. Bartley, J. Sperling, C. Silberhorn, and W. Vogel,
    	\textit{Incomplete Detection of Nonclassical Phase-Space Distributions},
    	\href{https://doi.org/10.1103/PhysRevLett.120.063607}{Phys. Rev. Lett. \textbf{120}, 063607 (2018)}.
    \bibitem{BBFPT19}
        M. Bouillard, G. Boucher, J. Ferrer Ortas, B. Pointard, and R. Tualle-Brouri,
        \textit{Quantum Storage of Single-Photon and Two-Photon Fock States with an All-Optical Quantum Memory}
        \href{https://doi.org/10.1103/PhysRevLett.122.210501}{Phys. Rev. Lett. \textbf{122}, 210501 (2019)}.
    \bibitem{SPBTEWLNLGVASW20}
    	J. Sperling, D. S. Phillips, J. F. F Bulmer, G. S. Thekkadath, A. Eckstein, T. A. W. Wolterink, J. Lugani, S. W. Nam, A. Lita, T. Gerrits, W. Vogel, G. S. Agarwal, C. Silberhorn, and I. A. Walmsley,
    	\textit{Detector-Agnostic Phase-Space Distributions},
    	\href{https://doi.org/10.1103/PhysRevLett.124.013605}{Phys. Rev. Lett. \textbf{124}, 013605 (2020)}.
	\bibitem{LHABMS01}
    	A. I. Lvovsky, H. Hansen, T. Aichele, O. Benson, J. Mlynek, and S. Schiller,
    	\textit{Quantum State Reconstruction of the Single-Photon Fock State},
	    \href{https://doi.org/10.1103/PhysRevLett.87.050402}{Phys. Rev. Lett. \textbf{87}, 050402 (2001)}.
	\bibitem{AT92}
		G. S. Agarwal and K. Tara, 
		\textit{Nonclassical character of states exhibiting no squeezing or sub-Poissonian statistics}, 
		\href{https://doi.org/10.1103/PhysRevA.46.485}{Phys. Rev. A \textbf{46} 485 (1992)}.
    \bibitem{ZPB07}
	    A. Zavatta, V. Parigi, and M. Bellini,
	    \textit{Experimental nonclassicality of single-photon-added thermal light states},
	    \href{https://doi.org/10.1103/PhysRevA.75.052106}{Phys. Rev. A \textbf{75}, 052106 (2007)}.
    \bibitem{BZ10}
        M. Bellini, and A. Zavatta,
        \textit{Manipulating Light States by Single-Photon Addition and Subtraction},
        \href{https://doi.org/10.1016/B978-0-444-53705-8.00002-3}{Progress in Optics
        \textbf{55}, 41 (2010)}.
    \bibitem{KV14}
	    B. K\"uhn and W. Vogel
	    \textit{Visualizing nonclassical effects in phase space},
	    \href{https://doi.org/10.1103/PhysRevA.90.033821}{Phys. Rev. A \textbf{90}, 033821 (2014).}
	\bibitem{BA20}
		M. Bohmann and E. Agudelo,
		\textit{Phase-space inequalities beyond negativities},
		\href{https://doi.org/10.1103/PhysRevLett.124.133601}{Phys. Rev. Lett. \textbf{124}, 133601 (2020)}.
	\bibitem{BAS20}
		M. Bohmann, E. Agudelo, and J. Sperling,
		\textit{Probing nonclassicality with matrices of phase-space distributions},
		\href{https://doi.org/10.22331/q-2020-10-15-343}{Quantum \textbf{4}, 343 (2020)}.
	\bibitem{PLN20}
	    J. Park, J. Lee, and H. Nha,
	    \textit{Verifying nonclassicality beyond negativity in phase space},
	    \href{https://arxiv.org/abs/2005.05739}{arXiv:2005.05739}.
	\bibitem{Z01}
	    W. H. Zurek,
	    \textit{Sub-Planck structure in phase space and its relevance for quantum decoherence},
	    \href{https://doi.org/10.1038/35089017}{Nature (London) \textbf{412}, 712 (2001)}.
	\bibitem{CG69}
	    K. E. Cahill and R. J. Glauber,
	    \textit{Ordered Expansions in Boson Amplitude Operators},
	    \href{https://doi.org/10.1103/PhysRev.177.1857}{Phys. Rev. \textbf{177}, 1857 (1969)}.
    \bibitem{A65}    
        F. T. Arecchi,
        \textit{Measurement of the statistical distribution of gaussian and laser sources},
        \href{https://doi.org/10.1103/PhysRevLett.15.912}{Phys. Rev. Lett. \textbf{15}, 912 (1965)}.
    \bibitem{HRFJ04} 
        Z. Hradil, J. \v{R}eh\'{a}\v{c}ek, J. Fiur\'{a}\v{s}ek, and M. Je\v{z}ek,
        \textit{Maximum-Likelihood Methodsin Quantum Mechanics},
        in \textit{Quantum State Estimation}, edited by M. Paris and J. \v{R}eh\'{a}\v{c}ek,
        \href{https://www.springer.com/gp/book/9783540223290#aboutBook}{Lecture Notes in Physics Vol. \textbf{649} (Springer, Berlin, 2004), pp.  59-112}.
    \bibitem{L04}
        A.I. Lvovsky,
        \textit{Iterative maximum-likelihood reconstruction in quantum homodyne tomography},
        \href{https://doi.org/10.1088/1464-4266/6/6/014}{J. Opt. B Quantum Semiclassical Opt.  \textbf{6}, S556 (2004)}.
    \bibitem{M79}
        L. Mandel, 
		\textit{Sub-Poissonian photon statistics in resonance fluorescence},
		\href{https://doi.org/10.1364/OL.4.000205}{Opt. Lett. \textbf{4}, 205 (1979)}.
    \bibitem{BCBZ20}
        N. Biagi, L. S. Costanzo, M. Bellini, and A. Zavatta,
        \textit{Entangling Macroscopic Light States by Delocalized Photon Addition},
        \href{https://doi.org/10.1103/PhysRevLett.124.033604}{Phys. Rev. Lett.
        \textbf{124}, 033604 (2020)}.
	\bibitem{GFPMTMGSD20}
		A. Grimm, N. E. Frattini, S. Puri, S. O. Mundhada, S. Touzard, M. Mirrahimi, S. M. Girvin, S. Shankar, M. H. Devoret,
		\textit{Stabilization and operation of a Kerr-cat qubit},
		\href{https://doi.org/10.1038/s41586-020-2587-z}{Nature (London) \textbf{584}, 205 (2020)}.
    \bibitem{MLLV19}
        J. T. Muhonen, G. R. La Gala, R. Leijssen, and E. Verhagen,
        \textit{State Preparation and Tomography of a Nanomechanical Resonator with Fast Light Pulses},
        \href{https://doi.org/10.1103/PhysRevLett.123.113601}{Phys. Rev. Lett. \textbf{123}, 113601 (2019)}.
    \bibitem{FH20}
        C. Fl\"uhmann and J. P. Home,
        \textit{Direct Characteristic-Function Tomography of Quantum States of the Trapped-Ion Motional Oscillator},
        \href{https://doi.org/10.1103/PhysRevLett.125.043602}{Phys. Rev. Lett. \textbf{125}, 043602 (2020)}.

\end{thebibliography}
\end{document}